# A method of combining traffic classification and traffic prediction based on machine learning in wireless networks

Wang Luming, Yang Mao*, Li Bo, Yan Zhongjiang

(School of Electronics and Information, Northwestern Polytechnical University, Xi'an 710072, China)

**Abstract**：With the increasing number of service types of wireless network and the increasingly obvious differentiation of quality of service (QoS) requirements, the traffic flow classification and traffic prediction technology are of great significance for wireless network to provide differentiated QoS guarantee. At present, the machine learning methods attract widespread attentions in the tuples based traffic flow classification as well as the time series based traffic flow prediction. However, most of the existing studies divide the traffic flow classification and traffic prediction into two independent processes, which leads to inaccurate classification and prediction results. Therefore, this paper proposes a method of joint wireless network traffic classification and traffic prediction based on machine learning. First, building different predictors based on traffic categories, so that different types of classified traffic can use more appropriate predictors for traffic prediction according to their categories. Secondly, the prediction results of different types of predictors, as a posteriori feature, are fed back to the classifiers as input features to improve the accuracy of the classifiers. The experimental results show that the proposed method has improves both the accuracy of traffic classification and traffic prediction in wireless networks.

## I Introduction

At present, with the rapid development of network technology, the complexity of the network is getting higher and higher, and the number of users of the network is also increasing exponentially, which makes the network information flow explode and the network congestion become more and more serious [1]. According to Cisco's annual Internet report, the number of Internet users will increase from 3.9 billion in 2018 to 5.3 billion in 2023 [2]. In addition, the recent Nokia report introduced and discussed various network traffic trends in 2020, which showed that network traffic increased significantly due to the COVID-19 pandemic. According to the data of ABI Research, a global technical intelligence company, the global 5G mobile data service traffic will reach 1676EB in 2026, with a CAGR of 63% [3]. In addition, in recent years, the rise and prosperity of ultra-large bandwidth applications (such as virtual reality, ultra-high-definition video, etc.) and real-time applications (RTA) have put forward higher requirements and challenges for wireless networks [4].

Quality of service (QoS) or quality of experience (QoE) is an important goal of wireless networks [5]. QoS and QoE are directly related to business characteristics. If the specific type of current service arrival can be accurately obtained through traffic classification, and the traffic characteristics of the network in the future can be grasped by accurately predicting the traffic sequence, each layer of the network (such as network layer, media access control (MAC) layer, etc.) will be able to adopt corresponding methods to provide QoS and QoE assurance methods accurately and pertinently [6]. However, on the one hand, the network is layered, and the application layer where the service is located and the lower layer (such as the MAC layer) are not tightly coupled [7]. This makes the characteristics of the upper layer service and the indication or marking of the QoS requirements attenuate layer by layer with the layer-by-layer transmission of the data packet, which makes it difficult for the lower layer to grasp the traffic characteristics when it receives the data packet of the service it carries. With the ever-changing business types and increasingly differentiated demand, this problem will become more serious. Therefore, how to accurately obtain the business type and predict the future traffic characteristics of the business becomes particularly important.

In recent years, industry and academia have gradually focused on traffic classification and traffic prediction through machine learning [8]. For example, in view of the ubiquitous traffic encryption and encapsulation phenomenon, the literature [9] describes the challenges faced by the classification and prediction task, explores the potential of machine learning in dealing with class imbalance and generalization problems and the ability to extract knowledge and analyze data from encrypted traffic, and demonstrates the advantages of machine learning in business classification and traffic prediction. Document [10] compares the machine learning method with the traditional linear models such as the Autoregressive Integrated Moving Average model (ARIMA) and the Simple Moving Average (SMA) model in detail. The experiment shows that the root mean square error of the machine learning method in traffic prediction is much lower than the traditional linear method. Literature [11] compares the effects of various machine learning methods on traffic prediction and classification in detail, and finds that the effect of neural network on classification and prediction is significantly better than other machine learning methods.

However, most of the work on traffic classification and traffic prediction is to separate the two. For business classification, we usually only rely on some prior features such as tuples, packet length, IP address, port number, protocol, etc. [12]. By relying on these prior features, we can usually achieve a high classification accuracy. However, because we do not add a posterior feature to the classification feature, there is a certain space for improvement in the structure. For traffic prediction, the existing research usually only uses the simple window moving method[13] to conduct online learning on the traffic time series. This method is less reliable, time-consuming and requires a large number of training set samples, because a single learner is difficult to learn all the network traffic patterns.

Therefore, in view of the problem that the classification and prediction results are limited due to the separation of traffic classification and traffic prediction, this paper proposes a joint traffic classification and traffic prediction method for wireless network traffic flow based on machine learning, which can improve the accuracy of traffic classification and traffic prediction at the same time. The core idea of this scheme is as follows: firstly, different predictors are constructed based on categories, so that businesses can select more appropriate predictors for traffic prediction according to their categories; At the same time, the prediction error of different types of predictors is fed back to the classifier as a feature to improve the accuracy of the classifier as a posteriori feature. The simulation results show that the accuracy of classification and prediction has been significantly improved by using this method.

The structure of the rest of the paper is as follows: Chapter 2 introduces the traffic classification based on machine learning and the sliding window model of traffic prediction; The third chapter introduces the methods of joint traffic classification and traffic prediction; The fourth chapter gives the design and result analysis of the simulation experiment using this model; Finally, the fifth chapter summarizes the work of this paper and proposes future research activities.

# II Related work

2.1 Business classification based on machine learning

Traffic classification is an important part of traffic analysis. It aims to divide Internet traffic into predefined categories, such as voice (VO) traffic, video (VI) traffic, game (GM) traffic, etc; Or classify according to the specific application source of the business. At present, a large number of machine learning algorithms have been applied to traffic classification, among which the representative algorithm is neural network, which has good robustness and fault-tolerance, high computing speed, strong learning ability and high classification accuracy.

At present, traffic classification usually depends on tuples (source IP, destination IP, source port, destination port, communication protocol). This traffic classification method based on address has certain disadvantages. For example, when the source IP and destination IP of service traffic are intranet addresses, the classification of the traffic will be difficult, or when the LAN adopts Dynamic Host Configuration Protocol (DCHP), The IP address of the server may be reassigned many times [14]. At this time, the training process of machine learning needs to be carried out again to relearn the newly assigned IP address. When the frequency of IP address reassignment is high, the business classification task will face great pressure.

For this reason, the classification features extracted from the traffic capture software in this paper are: the number of packets arriving within 0.5s, the exchange frequency of the sending direction within 0.5s, and the respective proportion and protocol of the two sending directions of the internal address within 0.5s.

Therefore, when the IP address and port number features cannot be used, the number of classification features obtained by the MAC layer is very limited, which also leads to the reduction of the accuracy of the traffic classification results. Moreover, since no posterior feature is added to the classification feature, there is a certain space for improvement in its structure. Therefore, the error of traffic prediction can be considered as the classification feature to improve the accuracy of classification.

2.2 Traffic prediction based on machine learning

The network traffic prediction problem is usually defined as the time series prediction problem. This kind of time series prediction problem can be solved by machine learning. Generally, machine learning prediction method has better prediction effect than traditional linear prediction model [15].

The most classical traffic prediction usually adopts the sliding window method, that is, a sequence point is used as the output of several previous sequences and slides down like a window to form a set of training sets. The advantage of the sliding window method is that it has low time complexity and is easy to construct. It only needs to determine the size of the sliding window and obtain a period of time series as required. Relying on machine learning, the internal relations and laws of the data in the traffic time series can be discovered and applied to traffic prediction.

However, there are great defects in using sliding window method to predict traffic. As shown in Figure 1, the training set contains three types of packages: VO, VI and GM. Different types of packets have different statistical characteristics. If different types of packets appear in the same training set, the statistical characteristics of these packets will interfere with each other. In this figure, when using the training set composed of VO package, VI package and GM package to test the GM package, the prediction effect will not be ideal due to the influence of the statistical characteristics of other types of packages. Another method that is theoretically feasible is to expand the capacity of the training set so that all possible sequences can be learned. However, this method will have huge data set mining difficulties, and need to collect a large amount of data; At the same time, it also takes a lot of time to learn and calculate such a large amount of data.

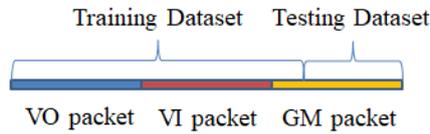

Figure 1 Sequence of different types of packets crossing

# III Method of joint classification and prediction of traffic flow

3.1 Overall design of the method

The model of business flow joint traffic classification and traffic prediction proposed in this paper is shown in Figure 2. The traffic flow sequence enters both the predictor group and the classifier group. The output of the predictor group includes the prediction results of the predictors corresponding to all possible categories. According to the current business category determined by the last classification result, the corresponding predictor prediction results will be used for practical applications; The prediction results of all predictors will be compared with the real value, and the error will be input into the classifier as a weight to assist classification. According to the output of the classifier and the error of the predictor, the classifier group is added with a certain weight to get the final classification result. When the traffic classification results are used for practical applications, they will also be input to the predictor to assist in traffic prediction and decide to output as the predictor for practical applications.

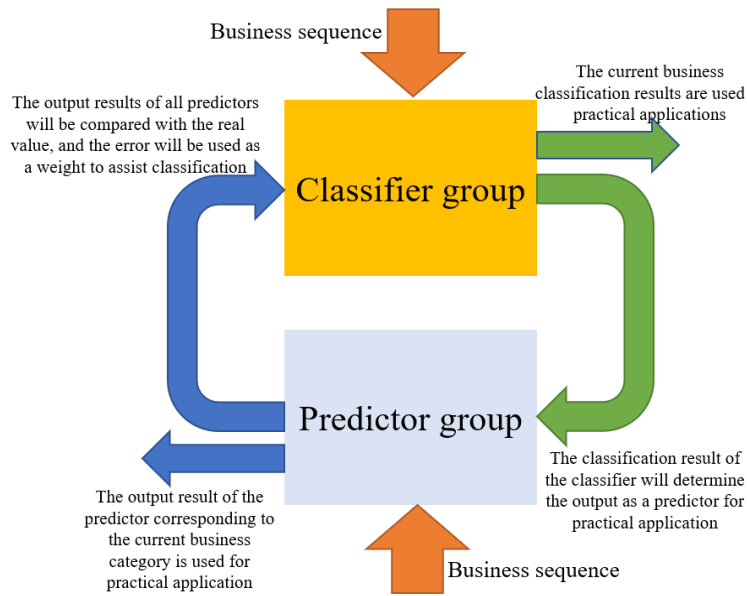

Figure 2 Model of joint classification and prediction of business flow

## 3.2 Classification-assisted prediction model

This paper adopts a method of using traffic classification results to assist traffic prediction, to achieve the purpose of classification assistance prediction. Firstly, the traffic sequence characteristics of different traffic flows are studied respectively, and different types of predictors are constructed. When the traffic flow arrives, first add the specific characteristics of the traffic flow and the error weight of the predictor, output the classification results, and then find the corresponding prediction model to predict according to the classification results. The specific process is shown in Figure 3.

For example, the predictor 1 learns the sequence characteristics of GM traffic; Predictor 2 learns the sequence characteristics of VO traffic, and predictor 3 learns the sequence characteristics of VI traffic. When the GM traffic arrives, the classifier is used to classify it first. According to the result, the predictor 1 corresponding to category 1 will be selected for prediction. Because predictor 1 learns the sequence characteristics of GM traffic, its prediction effect is better in theory.

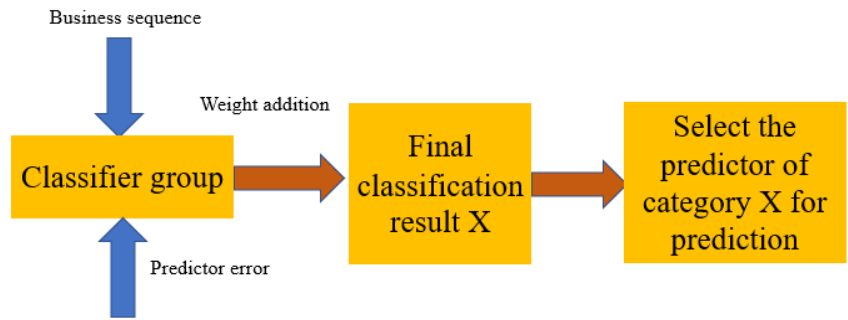

Figure 3 Classification-assisted prediction model

According to this method, the pressure of training set can be significantly reduced, and more traffic sequence patterns can be learned with less computational cost, and the time complexity and computational complexity can be reduced.

## 3.2 Model for predicting auxiliary classification

Traditional classification feature selection is mostly based on prior features, such as tuple, packet spacing, and so on. If prediction can be added to assist classification and used as a posteriori feature, the accuracy of classification will be improved in theory.

This paper proposes a scheme to use the prediction results of different traffic predictors to assist in traffic classification. The model is shown in Figure 4.

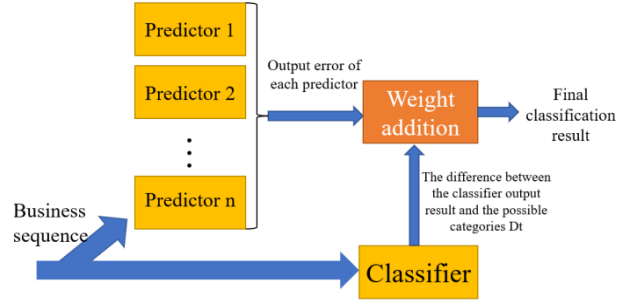

Figure 4 Model of Prediction-assisted Classification

Specifically, let the set of numerical tags corresponding to all categories in the training set be $X=\{X_1, X_2... X_n\}^T$. When the incoming traffic is sent into the classifier, a numerical solution $X_t$ will be obtained. Unlike the usual numerical dissociation value set X, where the nearest one will become the classification result of the business, this paper calculates the distance between $X_t$ and all elements in X to form a column vector $D_t$. Among them,

$$D_t = \left| X \begin{pmatrix} 1 \\ 1 \\ \vdots \\ 1 \end{pmatrix}_{n \times 1} - X_t \right|  \quad (1)$$

At the same time, different predictors are built according to the service category, and the traffic sequence corresponding to the service is imported into all predictors. The interval time of traffic prediction should be less than the interval time of traffic classification. Therefore, several traffic predictions will be made before traffic classification, and the gap between the real value and the predicted value can be obtained before the traffic classification, so the prediction error column vector $D_p$ of different types of predictors can be obtained.

Add $D_p$ and $D_t$ according to a certain weight value to get the final distance between the business flow and all categories:

$$D_a = D_p + \alpha D_t \quad (2)$$

Among them, α is a constant value, representing the weight between the characteristic result and the prediction error result. Select the category corresponding to the minimum value in $D_a$, which is the final business classification result.

## IV  Simulation experiment and result analysis

4.1 Data set used in the experiment

The data set used in the experiment is all from the measured data of the campus network, which is captured by the specific software first, and then processed by the data before the experiment.

Table 1 Data collection results

| business type | application | software files |
|---|---|---|
| VO business | WeChat voice | 8 |
| | QQ voice | 9 |
| VI business | Douyu Live | 9 |
| | Tencent Conference | 5 |
| | Kuaishou live | 11 |
| GM business | Honor of Kings | 3 |
| | PUBG | 3 |

In order to construct various types of classifiers and predictors, the collector will use the specific functions of the specific software in the process of capturing traffic to ensure that the traffic data collected in a specific period of time belong to the same business type and come from the same application software. Therefore, the data set can be labeled according to the software functions used by the collector when capturing traffic. After capturing the traffic data, convert the file into txt format in the virtual machine for data analysis. The collection results are shown in Table 1. Each file contains the traffic captured within 5 minutes. Select the traffic sequence of the first 2 minutes as the training set, and the traffic sequence of the last 2 minutes as the test set.

4.2 Simulation scenario design and result analysis

In order to verify the conjecture that using a specific type of predictor to predict traffic will be more accurate, this experiment uses the following modeling. First, the data is divided according to the business type, that is, the possible classification results are VO business, VI business or GM business, and the predictor is also built according to the above three businesses. Secondly, the basis for determining the classification task is the average packet interval of arriving packets within 0.5s, the number of exchanges between the sending direction, the number of times the user sends the contract to the other party, and the type of agreement; Determine that the traffic sequence of the prediction task application is the total length of the data packets that arrive within every 0.1s. Thirdly, according to the advantages of neural network in machine learning, this experiment selects BP neural network as the learner.

The experiment includes the comparison of the following three situations:

A. The strategy of assisting traffic prediction by business classification is adopted, and the classification results are correct. Use the same predictor as this category to predict the flow. That is, in the first layer, use the VO predictor to predict the VO flow, use the VI predictor to predict the VI flow, and use the GM predictor to predict the GM flow.

B. The strategy of using traffic classification to assist traffic prediction is adopted, but the classification result is wrong. Use a predictor different from this category to predict the flow, for example, use a GM type predictor to predict VO flow.

C. The strategy of auxiliary traffic prediction by business classification is not adopted. That is, integrate all traffic data together. The training set includes all kinds of traffic sequences, but the total length of the training set is consistent with the above situation. A single predictor learns the traffic sequence patterns of all business types and applications.

This experiment implements simulated RAM on a personal notebook (PC) with CPU (Intel Core i5-3210M, main frequency 2.5GHz). The root mean square error (RMSE) of the prediction results for scenarios A, B and C is shown in Table 2.

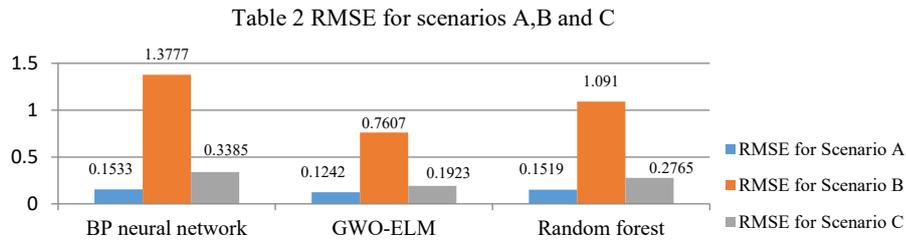

Table 2 RMSE for scenarios A,B and C

It can be seen from the table that scenario A, that is, using the prediction auxiliary classification method proposed in this paper, can control the RMSE to about 0.15 or less, that is, to achieve the prediction accuracy of about 85%. The RMSE of scenario C, which does not adopt the strategy of classification assisted prediction, is significantly higher than that of scenario A. Therefore, more accurate prediction results can be obtained by using the classification results and selecting the corresponding predictor. Because in Scenario C, the learner needs to learn not only the traffic sequence pattern of the current business type, but also the behavior pattern of other business types. However, the learner cannot distinguish different business types. Therefore, there will be interference between the traffic sequence patterns of different business types, reducing the accuracy of business prediction. However, the mode of selecting predictor based on category also depends on the accuracy of classification. When the classification result is wrong, the prediction result will lose its meaning. In Scenario B, RMSE is greater than 1,

In order to verify the auxiliary effect of traffic prediction results on traffic classification, three machine learning algorithms were selected in this experiment: the Extreme Learning Machine Algorithm (GWO-ELM) optimized based on Grey Wolf Algorithm, BP neural network, and decision tree. The first three minutes of data from the file were used as training sets for learning industry business types and traffic sequence patterns. The last two minutes of data from all files were used for testing, and business types were classified based on weight α. Calculate the average classification accuracy and observe the relationship between classification accuracy and weight α.

The relationship between the accuracy of traffic classification and α weight is shown in Figure 5.

In Figure 5, when α is 0, the corresponding accuracy represents not using traffic prediction to assist in traffic classification. Giving α a certain weight represents using the prediction assisted classification strategy mentioned in this article. The larger α, the greater the weight of the prediction error in classification. It can be seen that as α increases, the accuracy of traffic classification for the three types of machine learning shows a trend of first increasing and then decreasing. That is to say, giving traffic prediction error a certain weight in traffic classification can improve the accuracy of traffic classification. It is worth mentioning that the optimal value of α is related to machine learning algorithms, the characteristics of the dataset, and the calculation method of errors. It cannot be generalized. In the actual implementation process, it is first necessary to test the optimal value of α. However, it is worth affirming that using traffic prediction errors to assist in traffic classification can improve classification accuracy. The relationship between accuracy and α weight is shown in Figure 5.

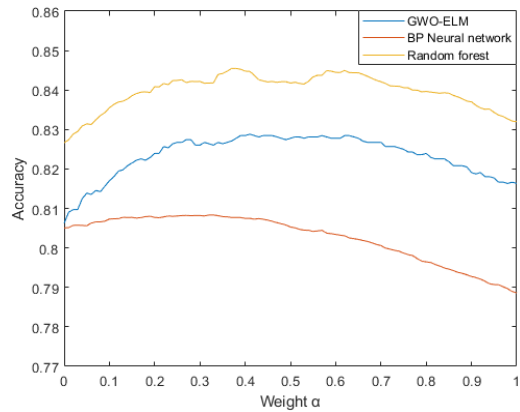

Figure 5 Accuracy and α Weight relationship

# V. Conclusion and later work

Aiming at the problem of combining traffic classification and traffic prediction, this paper proposes a method of combining traffic classification and traffic prediction for service flows. Different predictors are constructed based on categories, so that the traffic flow can select more appropriate predictors for traffic prediction according to their categories; At the same time, the prediction error of different types of predictors is fed back to the classifier as a feature to improve the accuracy of the classifier as a posteriori feature. The experimental results show that classification and prediction can promote each other. Using more accurate predictors for traffic classification can reduce RMSE by three times; Calculating the prediction error of different predictors and adding a certain weight to the classification can significantly reduce the classification error rate.

This paper discusses the combination of traffic flow classification and traffic prediction, and verifies that they can promote each other. Therefore, we can integrate the idea of traffic classification and traffic prediction into the MAC protocol in the later work, study the corresponding MAC guarantee scheme, and test the change of throughput on the wireless network simulation platform to make the research more in-depth and complete.

## ACKNOWLEDGEMENT

This work was supported in part by the National Natural Science Foundations of CHINA (Grant No. 61871322, No. 61771390, and No. 61771392).